\documentclass[letter,prapplied,
 amsmath,amssymb,
 reprint,
 author-year,%
superscriptaddress,longbibliography
]{revtex4-1}

\usepackage{natbib}
\usepackage[dvipsnames]{xcolor}
\usepackage{float}
\usepackage{graphicx}
\usepackage{tabularx}
\usepackage{epstopdf}
\AppendGraphicsExtensions{.tif}
\usepackage{amsmath,amsthm,amssymb,amsfonts}
\usepackage{dcolumn}
\usepackage{bm}
\usepackage{gensymb}
\usepackage[normalem]{ulem}

\newcolumntype{C}{>{\centering\arraybackslash}X} 

\usepackage{cleveref}
\crefname{equation}{Eq.}{Eqs.}
\Crefname{equation}{Equation}{Equations}
\crefname{figure}{Fig.}{Figs.}
\Crefname{figure}{Figure}{Figures}
\crefname{section}{Sect.}{Sects.}
\Crefname{section}{Section}{Sections}
\crefname{table}{Table}{Tables}
\crefname{appsec}{Appendix}{Appendices}
\graphicspath{{Figures/}}
\definecolor{kvkcolor}{rgb}{1,0.5,0.05}

\colorlet{Plamen}{blue}
\colorlet{P}{blue}
\colorlet{Kostya}{green}
\colorlet{K}{green}
\colorlet{Wen-Sen}{purple}
\colorlet{WS}{purple}
\colorlet{Misha}{red}
\colorlet{M}{red}
\colorlet{Tom}{OliveGreen}
\colorlet{T}{OliveGreen}

\begin{document}

\title{Granular aluminum meandered superinductors for quantum circuits}

\author{Plamen Kamenov}
\author{Wen-Sen Lu}
\author{Konstantin Kalashnikov}
\author{Thomas DiNapoli}
\affiliation{Department of Physics and Astronomy, Rutgers University, Piscataway, NJ}

\author{Matthew T. Bell}
\affiliation{Engineering Department, University of Massachusetts Boston, Boston, MA}

\author{Michael E. Gershenson}
\affiliation{Department of Physics and Astronomy, Rutgers University, Piscataway, NJ}

\date{\today}

\date{\today}

\begin{abstract}

We have designed superinductors made of strongly disordered superconductors for implementation in ``hybrid'' superconducting quantum circuits. The superinductors have been fabricated as meandered nanowires made of granular Aluminum films. Optimization of the device geometry enabled realization of superinductors with the inductance $\sim 1\, \mathrm{\mu H}$ and the self-resonance frequency over $\mathrm{3\, GHz}$. These compact superinductors are attractive for a wide range of applications, from superconducting circuits for quantum computing to microwave elements of cryogenic parametric amplifiers and kinetic-inductance photon detectors.

\end{abstract}

\pacs{Valid PACS appear here}
\keywords{Suggested keywords}
\maketitle

\section{Introduction}
Superinductors are non-dissipative elements with an impedance $Z$ much greater than the resistance quantum $R_Q=h/(2e)^2$  \cite{Manucharyan2012}. The realization of high impedance enables the enhancement of quantum fluctuations of phase in Josephson circuits at the expense of reduction of quantum fluctuations of charge. One of the important functions of superinductors in qubits such as fluxonium  \cite{Manucharyan2009} is the suppression of qubit dephasing induced by low-frequency charge noise. Superinductors are considered as prospective elements of traveling-wave parametric amplifiers (TWPA)\, \cite{Yamamoto2014,Eom2012,zhangKerr}, microwave kinetic inductance detectors \cite{Zmuidzinas2012}, tunable qubit couplers \cite{Peltonen2018}, and \textcolor{black}{superconducting nanowire} single-photon detectors\, \cite{gol'tsman,Voronov2011}.

Since the impedance of elements with conventional, “geometric” inductance is limited by the free-space impedance $Z_0  \approx 377 \, \mathrm{\Omega}$, superinductors must employ the inductance of superconductors related to the inertia of the Cooper pair condensate. Superinductors have been realized as chains of small Josephson junctions with high Josephson inductance $L_J$ \cite{Manucharyan2009,Bell2012,Weissl2015,DeGraaf2019} and long nanowires made of superconductors with a high kinetic inductance $L_K$ \cite{Hazard2018,Bylander2019, Grunhaupt2019}.

\textcolor{black}{Each implementation of superinductors has its limitations. In particular, an increase of the inductance of Josephson junctions by reducing the Josephson energy leads to undesirable increase of the phase-slip rate. The increase of kinetic inductance of nanowires by enhancement of disorder and associated rapid decrease of the superfluid stiffness near the disorder-driven superconductor-to-insulator transition (SIT) may result in emergence of sub-gap delocalized modes and stronger dissipation at microwave frequencies \cite{Feigelman2018}. Because of these limitations on increase of $L_k$, careful minimization of the stray capacitance $C$ of superinductors is required for the increase of their impedance $Z=\sqrt{L_k/C}$ and the self-resonance frequency $f_r=1/(2\pi\sqrt{L_k C})$, which is desirable for many superinductor applications (see, e.g. \cite{Santavicca2016}). }  

\textcolor{black}{Optimization of the superinductor design aims at increasing $f_r$ for a given impedance $Z$ dictated by applications. The importance of superinductor geometry has been often overlooked in optimization of high-impedance devices. Below we show that such meander parameters as the line spacing, overall size, and aspect ratio can significantly affect the total capacitance and, thus, the achievable impedance. In particular, we paid special attention to reducing the stray capacitances between different elements of the superinductor and between the superinductor and the ground.}

The aim of this paper is to optimize nanowire-based superinductors made of strongly disordered \textcolor{black}{superconductors, such as granular} Aluminum films. A combination of high kinetic inductance and relatively low microwave losses in the superconducting state makes these films attractive for implementation of superinductors \cite{Grunhaupt2018,Zhang2019}. Design optimization enabled us to fabricate superinductors with $L \textcolor{black}{>} 1\, \mathrm{\mu H}$ and the self-resonance frequency $\textcolor{black}{>} 3\,  \mathrm{GHz}$. We have also successfully integrated these superinductors in \textcolor{black}{“hybrid”} superconducting circuits containing conventional $\mathrm{Al-Al_2O_3-Al}$ Josephson junctions. 

\section{Simulations}

\begin{figure*}[t]
\centering
\includegraphics[scale=0.65]{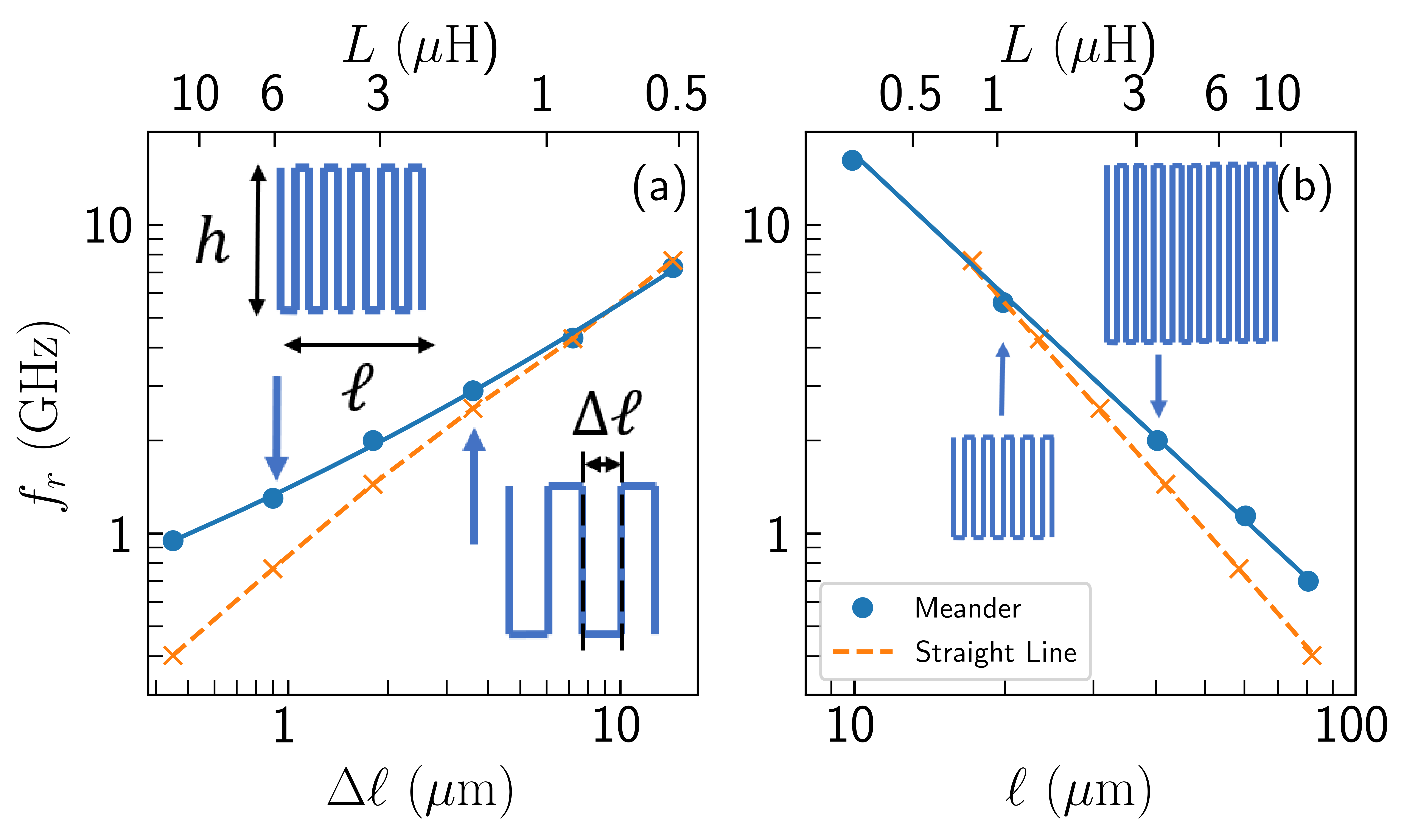}
\includegraphics[scale=0.385]{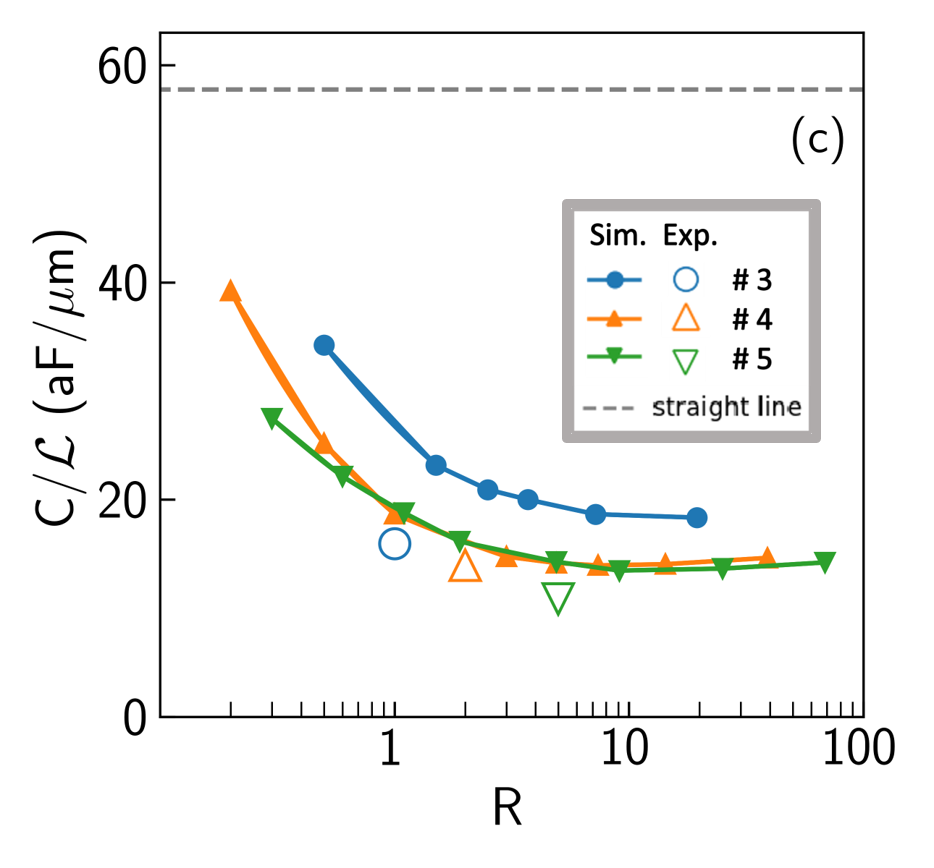}
\caption{(a) The simulated self-resonance frequency for the square-shaped meanders as a function of the meander period $\Delta \ell$ (blue dots). For all meanders, the side length $\ell=h = 40\, \mathrm{\mu m}$, the aspect ratio $R=\ell/h=1$, the nanowire width $W=0.3\, \mathrm{\mu m}$, \textcolor{black}{and the inductance per square $L_{\Box}=1nH$ close to the upper limit imposed by the SIT.} In comparison, we plot the self-resonance frequency for a straight nanowire with the total length equal to that for a corresponding meander (dashed orange line). (b) The simulated self-resonance frequency (blue dots) for square-shaped meanders as a function of the side length $h=\ell$. For all meanders $\Delta \ell=1.5\, \mathrm{\mu m}$, $W=0.3\, \mathrm{\mu m}$, \textcolor{black}{and $L_{\Box}=1nH$}.  The dashed orange line corresponds to the resonance frequency for a straight nanowire with the total length equal to that for a corresponding meander. \textcolor{black}{(c) Simulated dependences of the capacitance normalized by the total length $\mathcal{L}$ for meanders with different values of the aspect ratio $R$. The other parameters such as the total length, $\Delta \ell$, and $L_{\Box}$ match that for devices 3-5 (see Table 1). The dashed line corresponds to a straight-line inductor. The experimental values of $C/\mathcal{L}$ for devices 3-5 are shown as empty markers, the simulation results are shown as filled markers. Solid lines in panels a-c are guides to the eye.}}
\label{fig:fr_vs_dimensions}

\end{figure*}

We used Sonnet EM software \cite{Sonnet} to optimize the design of nanowire-based meandered superinductors, i.e. to increase the superinductor resonance frequency $f_r$ for a given total inductance and realistic nanowire dimensions. For the simulations in this section, we considered rectangular-shaped meandered superinductors, and adopted the width of the nanowire $W=0.3\, \mathrm{\mu m}$. \textcolor{black}{The self-resonance frequency simulations shown in Figs. 1a and 1b have been performed for $L_{\Box}=1\, \mathrm{nH}$, which is close to the upper limit of $L_{\Box}$ imposed by the SIT.} According to the Mattis-Bardeen theory, this value of $L_{\Box}$ can be realized for strongly disordered Al films with the normal-state sheet resistance $R_{\Box}\approx 1 \, \mathrm{k \Omega}$ \cite{Zhang2019}. \textcolor{black}{These simulations show the potential of superinductors made of granular Aluminum films. The self-resonance frequencies in Fig. 1c were calculated for the tested devices with $L_{\Box}$ within the range $0.04 - 0.23$ nH (parameters of these devices are listed in Table 1).} 

Meander optimization has been performed by varying the meander period $\Delta \ell$ and the aspect ratio $R=\ell/h$, where $\ell$ and $h$ are horizontal and vertical dimensions of a meander in \cref{fig:fr_vs_dimensions}, respectively. Note that the vertical dimension corresponds to the orientation of longer meander strips.

In simulations, it was assumed that the meanders were fabricated on a 0.5-$\mathrm{mm}$-thick Si substrate ($\epsilon_r=11.9$) with a negligibly thin native surface oxide, and the substrate was positioned on a conducting ground plane. The microwave currents in the meanders were induced by coupling to a microstrip line not shown in the insets in \cref{fig:fr_vs_dimensions}. The relative positioning of the meander and the microstrip line affects the coupling strength but not the value of $f_r$. 
The superinductor resonance frequencies have been identified as the frequencies of the resonance dips in the microwave power transmitted along the microstrip line. \textcolor{black}{For single meanders or pair of meanders (devices 1,2 in Table 1), the lowest-frequency resonance corresponds to a half-wavelength mode. For the qubit-based devices 3-4, the lowest resonance frequency corresponds to a full-wavelength mode due to the closed-loop boundary conditions discussed below.}

In \cref{fig:fr_vs_dimensions}(a) we compare the resonance frequencies for the square-shaped meanders ($\ell=h=40\, \mathrm{\mu m}$) with different values of $\Delta \ell$ and the straight nanowires with the same total length. We consider the meander to be composed of $l/\Delta l$ units of one vertical segment of length $h$ and one horizontal segment of length $\Delta l$ and one additional segment of length $h$ for a total length $\mathcal{L}=\ell(h+\Delta \ell)/\Delta \ell+h$. For meanders with closely spaced strips, the stray capacitance becomes significantly smaller than that for straight nanowires, which results in expected (and desired) increase of $f_r$.

\Cref{fig:fr_vs_dimensions}(b) shows scaling of $f_r$ with the dimensions of square-shaped meanders at a fixed $\Delta \ell=1.5\, \mathrm{\mu m}$. The dependence demonstrates that an increase of $f_r$ (and decrease of the stray capacitance) relative to that of a straight nanowire is more pronounced for larger meanders.

Interestingly, the square-shaped meanders do not maximize $f_r$. \textcolor{black}{\Cref{fig:fr_vs_dimensions}(c) shows that the dependences $\frac{C}{\mathcal{L}}(R)$, simulated for the tested devices (0.5 mm-thick Si substrates, $W=0.3\, \mathrm{\mu m}$ and $\Delta \ell=0.45\, \mathrm{\mu m}$), exhibit a minimum at the aspect ratio $R=\ell/h\simeq 5$.} The non-monotonic character of these dependences stems from the interplay of two competing factors. On the one hand, the capacitance to the ground of a solid rectangular conductor with a fixed area is minimized for its square shape. On the other hand, the inter-strip capacitance, which starts playing a significant role near the meander self-resonance, increases with the length of the strips (i.e. $h$). As a result of this competition, the minimum of stray capacitance occurs at $\ell>h$ or $R>1$. For the parameters used in simulations, the maximum self-resonance frequency of a meander is approximately twice as large as $f_r$ for a straight nanowire of the same total length and inductance \textcolor{black}{(see Table 1). The position of the minimum of $\frac{C}{\mathcal{L}}(R)$ depends on specific values of $\Delta \ell$ and $W$}. \textcolor{black}{In \Cref{fig:fr_vs_dimensions}(c), we plot the capacitance normalized by the total length for meanders with parameters listed in Table 1. For each $\mathcal{L}$, there is an optimal aspect ratio at which the self-resonant frequency significantly exceeds that for the straight-line inductor.}

\section{Fabrication}

\begin{figure}
    \centering
    \includegraphics[scale=.7]{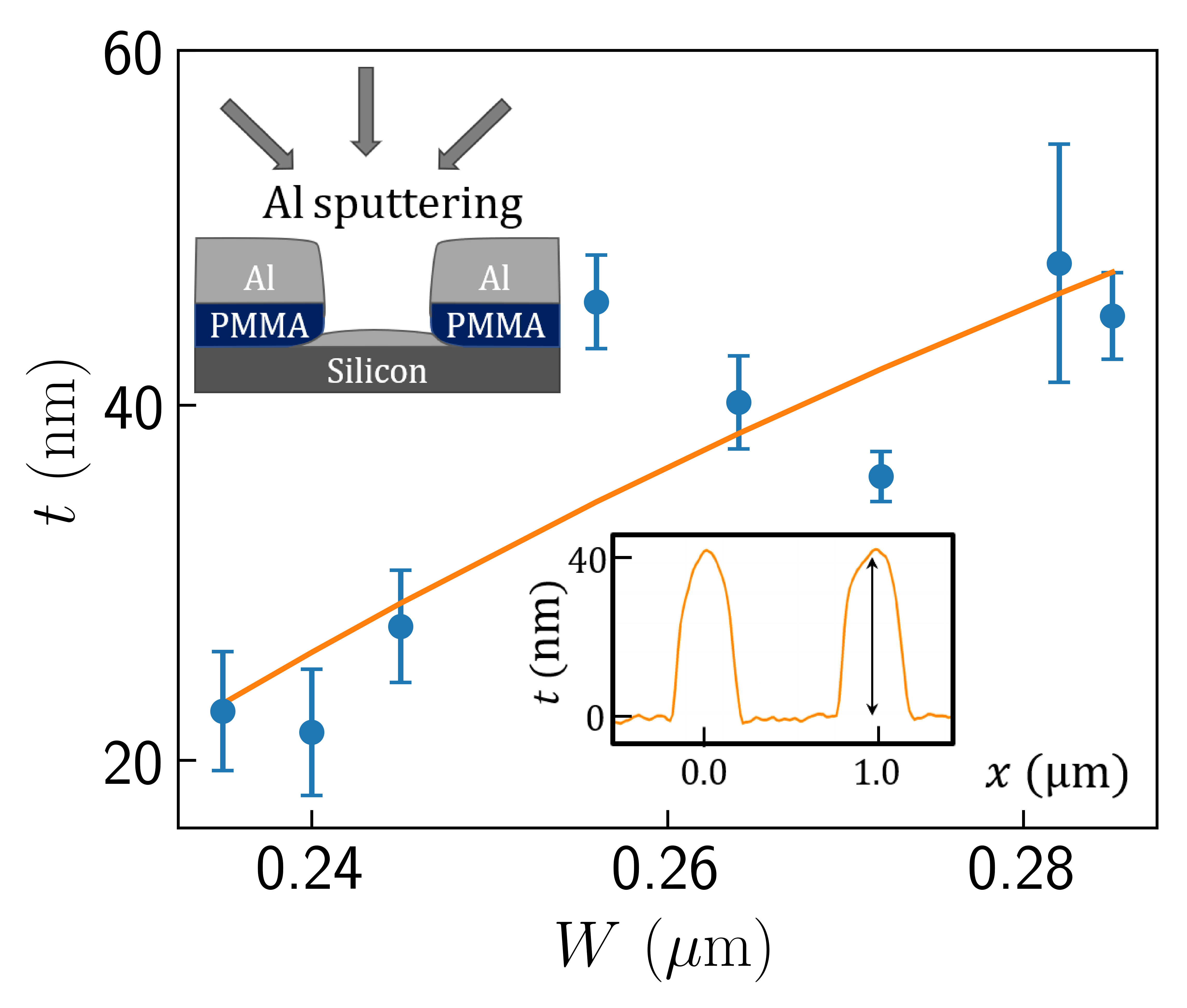}
    \caption{The dependence of the film thickness $t$ on the resist trench width $W$ measured by an AFM. The thickness of an un-patterned film in this deposition was 160 $\mathrm{nm}$. The top inset schematically illustrates the process of Al sputtering into a trench in PMMA resist. The bottom inset shows an AFM scan in the direction perpendicular to the meander strips. Note the dome-shaped profile of the nanowire cross-section due to shadowing. This profile was taken into account in calculations of the film resistivity.}
    \label{fig:film_tickness}
\end{figure}


We used disordered Aluminum films ($\mathrm{AlO_x}$) for fabrication of meandered superinductors. The films were deposited onto Silicon substrates by RF magnetron sputtering of a pure Al target in an Ar and $\mathrm{O_2}$ atmosphere \cite{Zhang2019}. The sheet resistance $R_{\Box}$ of these films, and, thus, their kinetic inductance in the superconducting state, $L_K=\hbar R_{\Box}/(\pi\times1.76 k_B T_C) $, can be tuned over a wide range by varying the deposition rate and the partial pressure of Oxygen in the deposition chamber. 
It is worth noting that the film thickness measured by atomic force microscopy (AFM) was approximately three times greater than the thickness measured by a quartz crystal monitor programmed with standard parameters for Al or $\mathrm{AlO_x}$ films. We speculate that this discrepancy is due to a low density of granular Al films sputtered in $\mathrm{O_2}$ atmosphere. 

The films were deposited on a substrate with a resist mask patterned by electron-beam lithography, and the device fabrication was completed by lift-off. In this fabrication process, thin $\mathrm{AlO_x}$ films were deposited in sub-$\mathrm{\mu m}$-wide resist \textcolor{black}{“trenches”} which were narrower than the liftoff mask thickness $\sim 0.5\, \mathrm{\mu m}$ (see the upper inset in \cref{fig:film_tickness}). Due to the “shadowing” effect, the deposition rate at the bottom of a trench is lower than for the films deposited on un-patterned substrates. Figure 2 shows that the meander thickness monotonically decreases with decreasing the nanowire width. The resistivity of $\mathrm{AlO_x}$ nanowires fabricated on the same chip did not significantly depend on the trench width.

For fabrication of “hybrid” superconducting circuits that combine high-$L_K$ $\mathrm{AlO_x}$ films with conventional $\mathrm{Al-Al_2O_3-Al}$ Josephson junctions, we used two deposition systems: the aforementioned sputtering system for fabrication of meandered superinductors and an electron-gun deposition system for fabrication of the Al-based tunnel junctions. The latter system was equipped with an ion gun which we used to remove thin ($\sim 4\, \mathrm{nm}$) oxide from the surface of $\mathrm{AlO_x}$ films prior to evaporation of pure Al films, to ensure a superconducting contact between the films.



\section{Measurements}

\begin{table}[t]
\centering
\caption{Parameters of several meandered superinductors in “hybrid” devices.
Each device contained two meanders which flanked either a single Josephson junction (JJ) (see the inset in \cref{fig:JJ_exp}) or a Cooper-pair-box qubit (see the inset in \cref{fig:qubit_exp}). The in-plane dimensions ($\ell \times h$) are given for each meander. The values of the total inductance $L$ and the self-resonance frequency $f_r$ correspond to pairs of meanders. The impedance was calculated as $Z=2\pi f_r\cdot L$.}

\begin{tabularx}{\columnwidth}{CCcCCCCC|C}
\hline
\hline

$\#$ & type& $	\ell \times h,$&	$L$,	&$L_{\Box},$&	$f_r^{exp}$,&	$f_r^{sim}$,  &	$Z$, & $f_{line}^{sim}$ \\

 & & $\mu m^2$ & $\mu H$ & $nH$ & GHz & GHz & $\mathrm{k\Omega}$ & GHz 
\\
\hline
1 &	JJ &	$20\times 20$&	1.32&	0.22&	3.30&	3.5 &	27.5 & 1.47 \\

2 &	JJ &	$20\times 20$&	1.15&	0.19&	3.16&	3.8 &	22.9 & 1.58\\
3 &	qubit &	$10\times 10$&	0.19&	0.14&	12.6&	13.8&	10.1 & 6.95 \\

4 & qubit & $20 \times 10$ & 0.11 & 0.04 & 12.5 & 12.0 & 8.6 & 6.61 \\

5 & qubit & $25 \times 5$ & 0.50 & 0.23 & 6.2  & 6.18  & 19.5 & 3.96 \\

\hline
\hline
\end{tabularx}
\label{table:sample_summary}
\end{table}

\begin{figure}[t!]
    \centering
    \includegraphics[width=\columnwidth]{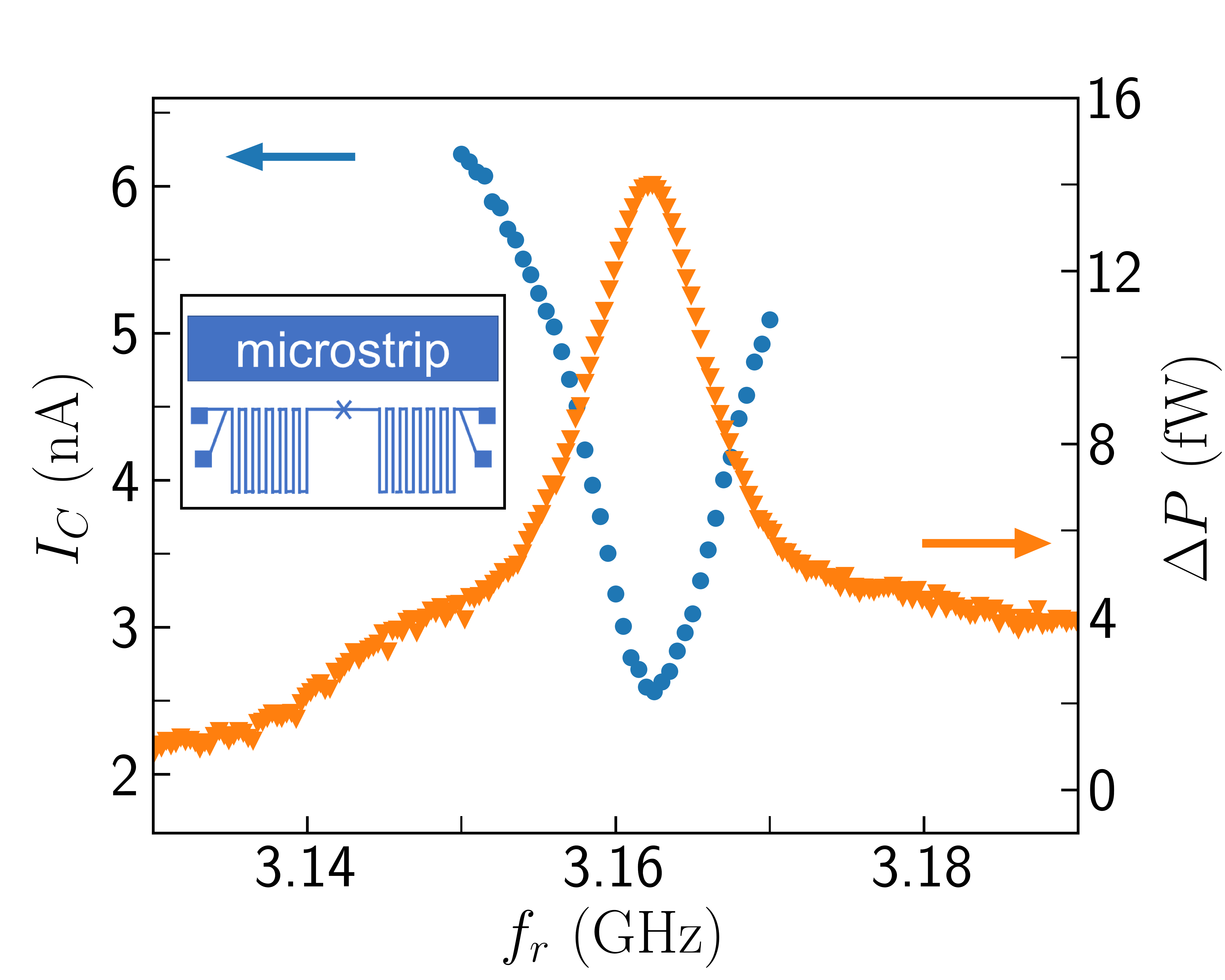}
    \caption{The frequency dependences of the critical current $I_C $ (blue dots) and the absorbed MW power $\Delta P \equiv P \mathrm{(3.1 \,GHz)}- P(f)$ (orange triangles) measured for device 1 at $T=20\,mK$. The inset shows schematically the design of devices 1 and 2.}
    \label{fig:JJ_exp}
\end{figure}

\begin{figure}[h]
    \includegraphics[width=\columnwidth]{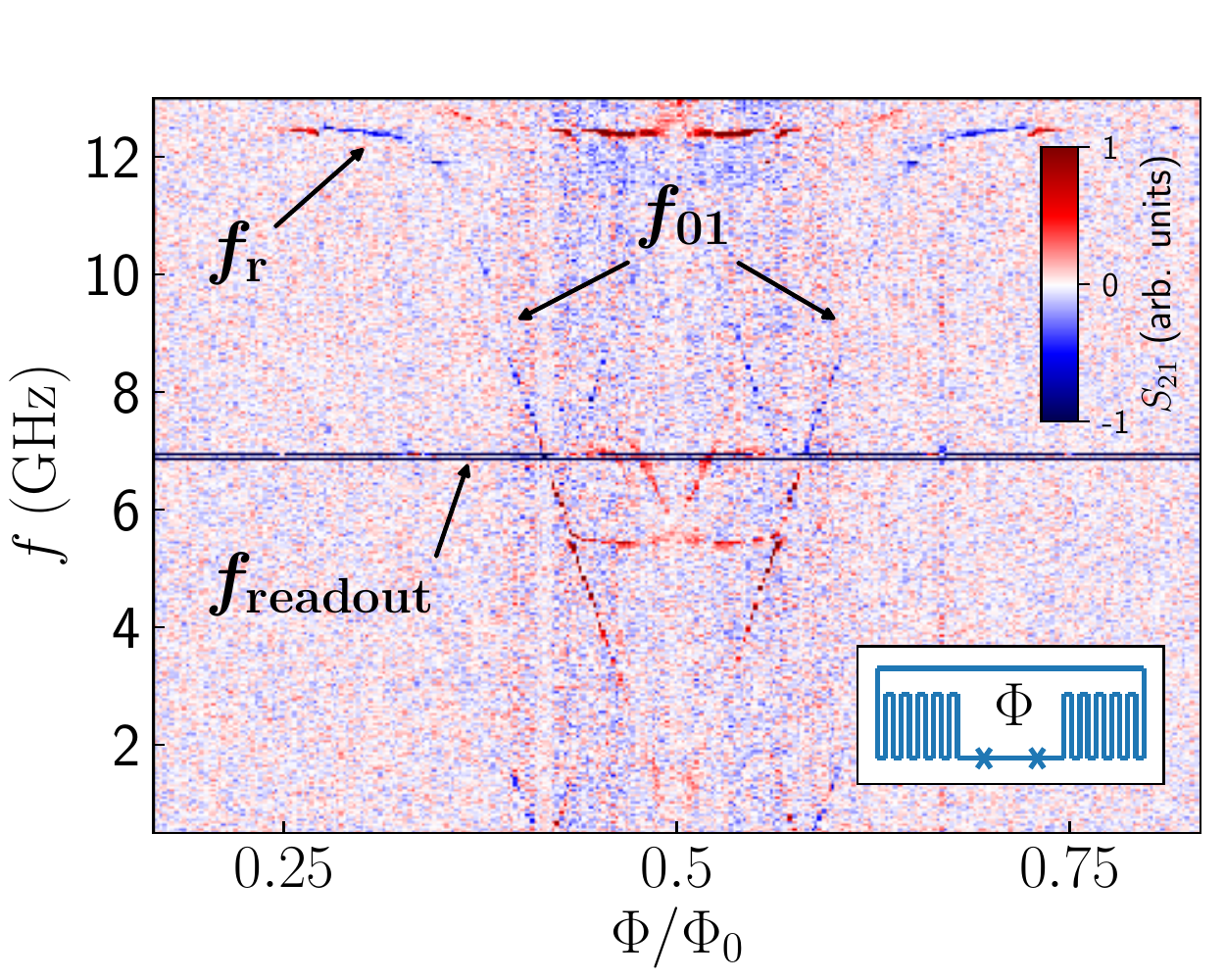}
    \caption{Microwave spectroscopy of bifluxon qubits (device 4). The features around half flux frustration and $f=5.5\, \mathrm{GHz}$ are due to multi-photon processes. The inset shows schematically the design of devices 3 and 4.}
    \label{fig:qubit_exp}
\end{figure}

To characterize the microwave properties of meandered superinductors, we tested two types of “hybrid” devices that contained both superinductors and Josephson junctions. We fabricated meanders from $\mathrm{AlO_x}$ films with $L_{\Box}=0.04-0.22\, \mathrm{nH}$ (the normal-state sheet resistance $R_{\Box}\simeq 50-300\,\Omega$). The nominal nanowire width $W=0.3\, \mathrm{\mu m}$ and the period $\Delta \ell=0.45\, \mathrm{\mu m}$ were the same for all meanders. The parameters of devices referenced below are listed in \cref{table:sample_summary}. In devices 1 and 2, two identical superinductors flanked a single Josephson junction (JJ) with an area of $A\sim 0.04\, \mathrm{\mu m}^2$
\textcolor{black}{(see the inset in \cref{fig:JJ_exp})}. The contact pads were used for the four-probe DC measurements. In devices \textcolor{black}{3-5}, the single JJ was replaced by a Cooper pair box (CPB) (see the inset in \cref{fig:qubit_exp}). The CPB and meanders formed a superconducting loop coupled to the read-out $LC$ resonator \cite{Bell2016}. The latter devices have been designed as prototypes of bifluxon qubits; similar devices with different implementations of superinductors have been studied in Refs. \cite{Bell2016,DeGraaf2018,bifluxon}.

In experiments with devices 1 and 2, microwave (MW) currents in the meanders were induced by coupling to a $100$-$\mathrm{\mu m}$-wide microstrip line that was positioned at a distance of $20 \, \mathrm{\mu m}$ from the meander along its horizontal dimension. At a fixed input MW power, we simultaneously measured the transmitted power $P\propto |S_{21}|^2$ in the line and the current-voltage characteristics (IVCs) of the devices as a function of the MW  frequency.
\Cref{fig:JJ_exp} shows a sharp resonance absorption of the MW power in device 1 near the resonance frequency $f_r \approx 3.16\; \mathrm{GHz}$ and strong suppression of the device's critical current at this frequency. Note that the critical current $I_C$ of the device is controlled by the JJ whose $I_C$ was smaller than that for the superinductor by three orders of magnitude. The observed suppression of $I_C$ was caused by an increase of the MW currents flowing through the JJ due to the resonant drop in magnitude of the superinductor's impedance at $f\sim f_r$. The total inductance of \textcolor{black}{these devices} was calculated using the low-temperature limit of the Mattis-Bardeen theory; the input parameters were the normal-state resistance $R_{\Box}$ and the critical temperature $T_C$.

In the experiments with devices \textcolor{black}{3-5} (bifluxon qubits), we used standard two-tone spectroscopic measurements \cite{Blais2004}, which provided both the self-resonance frequency of a superinductor and its total inductance extracted from the magnetic flux dependence of the qubit spectrum (\cref{fig:qubit_exp}, see \cite{Bell2016,bifluxon} for details). In these experiments, the superinductor mode appears as a resonance independent of the flux in the device loop and the charge on the CPB island. The inductance of the meanders can be extracted from the slope of the qubit resonance frequency as a function of flux using

\begin{equation}
    \frac{\partial f_{01}(\Phi)}{\partial \Phi}=\pm \frac{1}{h} \frac{\Phi_0}{2L},
\end{equation}
where $h$ is Planck's constant and $\Phi_0$ is the flux quantum.

Since the DC resistive measurements could not be performed on the qubits, the extracted inductance was used as an input parameter in the simulations. The geometry of these devices - two meanders connected together to form a loop - affected the boundary conditions. As a result, the lowest-frequency mode corresponded to two current density nodes between the meanders, in contrast to the fundamental mode of devices 1 and 2 with the current density anti-node between the meanders \cite{Khalil2011}.

\cref{table:sample_summary} shows that the simulations agree well with the measured values of $f_r$. \textcolor{black}{In \cref{fig:fr_vs_dimensions}(c), we compare the experimental results obtained for the qubit-based devices with our simulations. The total capacitance of the meanders is significantly lower than the simulated capacitance of a straight-line inductor of the same total length.} One conclusion derived from this comparison is that the Mattis-Bardeen theory is reliable for calculating the kinetic inductance of $\mathrm{AlO_x}$ films with the sheet resistance up to \textcolor{black}{$0.3\, k\Omega$}. The impedance of the realized meandered superinductors, $Z=2\pi f_r \cdot L=10-30\, \mathrm{k\Omega}$, significantly exceeded the resistance quantum $R_Q=h/(2e)^2 \simeq 6.5\, \mathrm{k\Omega}$. 

\textcolor{black}{The experiments with the fluxon-parity-protected qubits (a.k.a. bifluxon qubits [21,23]) provide important information on dissipation in the nanowire-based superinductors. The decoherence mechanisms in bifluxon qubits with superinductors implemented as chains of Josephson junctions have be discussed in Ref. [23]. For the nanowire-based bifluxon qubits at the resonance frequency $f_{01}=2$ GHz, we have measured the decay time $T_1\simeq 6 \mu$s in the unprotected regime, which corresponds to the quality factor $Q=2\pi f_{01}\cdot T_1\simeq 10^5$. Similar dissipation in the unprotected regime was observed for the bifluxon qubits with the Josephson-junction-based superinductors. This suggests that, at least at this overall dissipation level, the nanowire-based superinductors do not introduce additional losses.}

\section{Conclusion}

We have shown that meandering of high-kinetic-inductance nanowires enables significant increase of the self-resonance frequency of these microwave devices intended as superinductors. We fabricated meanders with a kinetic inductance of $0.1\,\mathrm{\mu H}-1.3\,\mathrm{\mu H}$, and self-resonance frequency between $12\, \mathrm{GHz}$ and $3\, \mathrm{GHz}$, respectively. The on-chip footprint of these meanders with a total nanowire length $\mathcal{L}$ up to $900 \, \mathrm{\mu m}$ does not exceed $400\, \mathrm{\mu m}^2$. The optimal in-plane shape of meanders can significantly deviate from a square due to an interplay between the strip-to-strip and strip-to-ground capacitances.

The impedance of the realized meandered superinductors, $Z\simeq10-30\, \mathrm{k\Omega}$, is sufficiently high for a range of applications, such as the flux-based superconducting qubits and high-kinetic-inductance detectors. However, even higher impedances are required for significant enhancement of quantum phase fluctuations in the ``$0-\pi$'' qubits \cite{Preskill2013,Groszkowski}, and for the devices based on Bloch oscillations \cite{Averin1985,Kuzmin1991,Arndt2018}.  Further increase of the impedance of meandered superinductors can be achieved by using more disordered films (i.e. increasing $L_{\Box}$), reducing $W$ and $\Delta \ell$ (i.e. decreasing overall meander dimensions for a fixed total inductance), and using substrates with a lower relative permittivity $\epsilon_r$ \cite{Pechenezhskiy2019}. A significant decrease of the nanowire width may be limited, in addition to lithographic resolution, by several factors that require further studies. For example, the reproducibility of kinetic inductance may be affected by percolation effects in quasi-one-dimensional nanowires fabricated from strongly disordered (granular) superconductors. Also, interactions between the nanowire and two-level systems in the environment can induce fluctuations of the local density of the Cooper pair condensate, and, thus, the  kinetic inductance of very narrow nanowires \cite{Zhang2019, LeSueur2018}.

\begin{acknowledgments}
We would like to thank Zhenyuan Zhang and Xinyuan Lai for help with AFM measurements and  Chung-Tse Michael Wu for helpful discussion. The work  at  Rutgers  University  was  supported  by the NSF award DMR-1708954 and the ARO award W911NF-17-C-0024.  The work at the University of Massachusetts Boston was supported in part by a 2019 Google Faculty Research Award and NSF Awards No. ECCS-1608448, DUE-1723511, and DMR-1838979.
\end{acknowledgments}

%

\end{document}